   \documentclass[prl,aps,twocolumn,showpacs]{revtex4}
\usepackage[dvips]{graphicx}
\usepackage{dcolumn}
\newcommand{\be}{\begin{equation}}
\newcommand{\ee}{\end{equation}}
\newcommand{\bea}{\begin{eqnarray}}
\newcommand{\eea}{\end{eqnarray}}
\newcommand{\nn}{\nonumber}
\newcommand{\ds}{\displaystyle}
\begin{document}
\topmargin=-20mm

\title{On the Theory of Quantum Oscillations of the Elastic Moduli in Layered Conductors}

\author{ Nataliya A. Zimbovskaya and Joseph L. Birman}

\affiliation{Department of Physics, The City College of CUNY, New York, NY, 10031, USA}

\date{\today}

\begin{abstract}   
In this paper we study theoretically how the local geometry of the Fermi surface (FS) of a layered conductor can affect quantum oscillations in the thermodynamic observables. We introduce a concrete model of the FS of a layered conductor. The model permits us to analyze the characteristic features of quantum oscillatory phenomena in these materials which occur due to local anomalies of the Gaussian curvature of the FS. Our analysis takes into account strong interaction among quasiparticles and we study the effect of this interaction within the framework of Fermi-liquid theory. We show that singularities in the density of states of quasiparticles (DOS) on the FS which occur at low temperatures near the peaks of its oscillations in a strong magnetic field can be significantly strengthened when the FS of the layered conductor is locally flattened. This can lead to magnetic and lattice instabilities of a special kind which are considered in the final part of the work.
              \end{abstract}

              \pacs{71.10 Pm, 73.40 Hm, 73.20 Dx }
\maketitle

\section{I. Introduction}

Many superconducting materials with large critical parameters created in the last two decades are layered structures with metallic-type conductivity (e.g organic metals of the $\alpha$-(BEDT-TTF)$_2$MHg(SCH)$_4$ group). A characteristic feature  of these materials is a strong anisotropy of the conductivity in the nonsuperconducting state: the conductivity in the layer plane is much higher than in the direction normal to the layers. It is common to assume that an anisotropy in electrical conductivity is a manifestation of the quasi-two-dimensional nature of the energy spectrum of the charge carriers in layered conductors. It follows from experiments that the Fermi surface (FS) of such conductors is a system of weakly rippled cylinders (isolated or connected by links) whose axes are perpendicular to the layers. The experimental data (see, e.g. Refs. \cite{1,2,3,4,5,6,7,8,9}) support this assumption.

The FS of a conductor with a quasi-two-dimensional energy spectrum can be described by the following equation
  \be 
E{\bf (p)} = \sum_{k=0}^\infty E_k (p_x,p_y) \cos \frac{a k p_z}{\hbar}
  \ee
   where $ \bf p $  is the electron quasimomentum; $ E_k (p_x,p_y) $ are coefficients with dimensions of energy, $ p_z $ is the projection of the quasimomentum on the direction normal to the layers, and $ a $ is the distance between the layers. If we ignore the anisotropy of the energy spectrum in the layer plane, instead of (1) we can write the simpler equation:
  \be 
  E_p = \frac{p_\perp^2}{2 m_\perp} +
\sum_{k=1}^\infty E_k (p_\perp) \cos \frac{a k p_z}{\hbar},
   \ee
  where $ \bf p_\perp $ is the projection of the quasimomentum on the layer plane, and $ m_\perp $ is the effective mass corresponding to the motion of the quasiparticles in that plane. Equation (2) describes an axially symmetric open FS with the axis of symmetry $(z $ axis of the chosen coordinate system) directed along a normal to the layers. The profile of the longitudinal section of the FS corresponding to Eq. (2) is a periodic function of period $ 2 p_0 \equiv 2 \pi \hbar /a. $

Experimental data concerning quantum oscillations in various characteristics of metals were widely used as the instruments of study of their electron energy spectra. Quantum oscillatory phenomena were repeatedly observed in organic layered conductors (Refs. \cite{1,3,4} and \cite{10}). Quantum oscillations in thermodynamic observables in these materials can exhibit some characteristic properties which occur due to the quasi-2D character of the energy spectrum of charge carriers. For instance it was shown in Ref. \cite{11} that the magnetic susceptibility of the noninteracting 2D electron system with open FS can exhibit sharp maxima due to the orbital effect. However, a systematic theory of quantum oscillatory phenomena in organic metals which takes into account the quasi-2D character of their electron spectra and a strong correlation among the electrons is not well developed at present. In this paper we give such theoretical analysis and we apply our results to study the quantum oscillations of the thermodynamic characteristics, especially of the elastic moduli.

At first, we derive the expressions for the electron contributions to the elastic constants corresponding to the compression in the direction perpendicular to layers  $ (\lambda_0) $ and to the compression applied along the layers $ (\lambda)$ in the presence of external magnetic field $ \bf B. $ In the geometry where the $ ``z"$ axis is directed along the normal to the layers, and assuming that the system is axially symmetric our elastic constants $ \lambda_0  $ and $ \lambda $ correspond to the electron contribution to the elastic moduli $ c_{33} $ and $ c_{11} = c_{22} $ in Voigt notation. A possible instability of the lattice arises near the peaks of the oscillations of DOS in a quantizing magnetic field. Therefore we analyze these oscillations of DOS in detail, and we show that they can be strongly affected due to the interaction among the carriers. In the following section of the work we introduce a concrete model of the FS which permits us to analyze how the local geometry of the FS can strengthen the singularities in DOS which can give rise to the considered lattice instabilities.

\section{2. Main equations}

To simplify further calculations we do not consider the deformation interaction between the electrons and the lattice. Using the generally accepted assumption the effect of the electrons on the lattice arises due to a self-consistent electric field, which occurs under deformation. Also, the deformation of the lattice causes the occurence of an additional non-uniform magnetic field $ \bf b(r) .$ The presence of these fields leads to a redistribution of the electron density $ N. $ The local change in the electronic density $ \delta N \bf (r) $ equals:
   \bea 
  \delta N {\bf (r)} &
=& - \frac{\partial N}{\partial \zeta} e \varphi {\bf (r)} + 
\frac{\partial N}{\partial \bf B} {\bf b (r)}  \nn \\
 &\equiv &
-N_\zeta^* \left [e \varphi {\bf (r)} + 
\frac{\partial \zeta}{\partial \bf B} {\bf b (r)} \right ].
   \eea
  Here, $ e $ is the absolute value of the electron charge, $ \bf B $ is the external magnetic field.

The magnetic field $ \bf b(r) $ satisfies the equation:
   \bea 
  \mbox{curl  \bf b(r)}& =& 4 \pi\, \mbox{curl $\delta$\bf M(r)} \nn \\ &=&
4 \pi \, \mbox{curl} \left [\frac{\partial \bf M}{\partial \zeta} e \varphi {\bf (r)} + \frac{\partial \bf M}{\partial \bf B} {\bf b (r)} \right ] \mbox{div \bf b(r) } \nn \\
  & =& 0.
   \eea
   Here, $ \bf M $ is the magnetization vector; $ \zeta $ is the chemical potential of charge carriers; $ \varphi \bf (r) $ is the potential of the electric field, arising due to the deformation; $ N_\zeta $ is the ``bare" density of states (DOS) of quasiparticles on the Fermi surface of the layered conductor:
   \be 
  N_\zeta = - \sum_\nu \frac{\partial f_\nu}{\partial E_\nu} .
   \ee
  $ f_\nu $ is the Fermi distribution function for the quasiparticles with energies $ E_\nu. $

Within the framework of quantum phenomenological Fermi-liquid theory the quantity $ N_\zeta^* $ is the DOS of the quasiparticles renormalized due to the Fermi-liquid interaction among them (see Ref. \cite{13}):
      \be 
N_\zeta^* = - \sum \limits_{\nu \nu'}
\frac{f_\nu - f_{\nu'}}{E_\nu - E_{\nu'}} \,
n_{\nu \nu'}^* (- {\bf \mbox q})
n_{\nu' \nu} ( {\bf \mbox q}) |_{q = 0} .
                                      \ee
   and the renormalized operator  of electron density $ n_{\nu \nu'}^*\bf (-q) $ is connected with the ``bare"operator 
$ n_{\nu \nu'}\bf (-q) $ by the relation:
          \be 
n_{\nu \nu'}^* ({\bf -q}) = n_{\nu \nu'} ({\bf -q}) + \sum
\limits_{\nu_1 \nu_2} \frac{f_{\nu_1} - f_{\nu_2} }{E_{\nu_1} - E_{\nu_2}} F_{\nu \nu'}^{\nu_1 \nu_2} n_{\nu_1 \nu_2}^* ({\bf - q})
       \ee
   where $F_{\nu \nu'}^{\nu_1 \nu_2}$ are the matrix elements of the Fermi-liquid kernel [See Eq. (13) below].

The relations (3), (4) have to be complemented by the condition of electrical neutrality of the system:
   \be 
  \delta N {\bf (r)} + e N \,\mbox{div \bf u(r) } = 0,
  \ee  
  where $ \bf u(r) $ is the lattice displacement vector, $ N $ is the electron density. It follows from equations (3), (4) and (8) that:
  \be 
   \mbox{curl} \{(1 -4 \pi \chi) {\bf b (r)}\} = - 4 \pi \,
 \mbox{curl} \left \{\frac{\partial \bf M}{\partial \zeta} 
\frac{N}{N_\zeta^*} \mbox{div \bf u(r)} \right \}.
   \ee
  The set of simultaneous equations (3), (4) and (8) was first presented in Refs. \cite{14} and \cite{15}. It allows us to exclude $ \bf b (r) $ and to express the potential $ \varphi \bf (r) $ in terms of the lattice displacement vector. As a result we can derive the expression for the electron force $ \bf F (r) $ acting upon the lattice under its displacement by the vector $ \bf u(r) $. For the conductor with axially symmetric FS in the magnetic field $ \b B $ directed along the symmetry axis, the force $ \bf F (r) $ equals:
  \be 
 {\bf F(r)} = \lambda_0 {\bf b}_0 ({\bf b}_0 \nabla ( \nabla {\bf u(r)))} + 
\lambda [{\bf b}_0 \times [\nabla (\nabla {\bf u(r))} \times {\bf b}_0]].
  \ee
   Here ${\bf b}_0 $ is the unit vector directed along  $ \bf B$. We assume that the magnetic field $ \bf B $ is directed along the normal to the layers. Then $ {\bf b}_0 $ coincides with the normal vector to the layers $ \bf n. $ It follows from Eq. (10), that under longitudinal (parallel to  $ {\bf b}_0 $ or perpendicular to layers) direction of gradients of the deformation tensor the electron contribution to the corresponding elastic modulus is equal to $ \lambda_0, $ and under their direction across $ {\bf b}_0 $ (in the plane of layers) it equals $ \lambda. $ The elastic constants $ \lambda_0 $ and $ \lambda $ are described by the expressions:
   \bea 
  \lambda_0 &=& \frac{N^2}{N_\zeta^*},
  \\
  \lambda &=& \lambda_0 \left (1 + \frac{4 \pi \chi_\zeta}{1 - 4 \pi \chi_{||}} \right ).
  \eea
  Here, $ \chi_{||} $ is the longitudinal part of the magnetic susceptibility; $ \chi_\zeta = (\partial M_z/\partial \zeta)(\partial \zeta/\partial B). $ More careful treatment carried out before including the deformation interaction (see e.g. Ref. \cite{16}), do not give a qualitative change of results. If we will take into account such interaction, it will lead us to a replacement of the electron density $ N $ in the expression (11) for $ \lambda_0 $ by the quantity $ N (1+ a), $ where the value of the dimensionless constant $ a $ depends on the corresponding component of the tensor of deformation potential averaged over the FS. The constant $ N $ in Eq. (12) also has to be replaced by $ N (1 + a^* ) $ where $ a^* $ is a dimensionless constant depending on the averaged over the FS deformation potential.

As follows from Eq. (11), in our case, when we neglect the deformation interaction the quantity $ \lambda_0 $ coincides with the compression modulus of the electron liquid. The structure of the quantity $ \lambda $ is more complicated. Besides the contribution,  connected  with the electron compression itself, $ \lambda $ also contains the contribution arising due  to the magnetostriction (the second term in the expression enclosed in the brackets). It is shown below that at low temperatures when $ \theta \ll 1 \ (\theta = 2 \pi^2 T/\hbar \Omega; \ T $ is the temperature expressed in energy units, $ \Omega $ is the cyclotron frequency of the charge carriers) there is a significant distinction between the oscillating corrections to the elastic moduli $ \lambda_0 $ and $ \lambda $ arising due to the magnetostriction. In the considered case $ \bf (B || n)$ the velocity of the longitudinal sound, propagating along the normal to the layers is proportional to the square root of $ \lambda_0, $ and when it propagates in the plane of layers -- to the square root of $ \lambda. $ Therefore, the distinction between $ \lambda_0 $ and $ \lambda $ due to the magnetostriction will be displayed in the quantum oscillations of the velocity of sound. Above the low temperature range $ (\theta \sim 1) $ the contribution connected with the magnetostriction becomes negligible and the distinction between $ \lambda_0 $ and $ \lambda $ is smoothed.

\section{3. low temperature quantum oscillations of the DOS}

Here, we calculate the oscillating corrections to the renormalized DOS  $N_\zeta^* $ which is defined by Eq. (6). The calculation is carried out for an axially symmetric FS of an arbitrary shape and this needs a special consideration because the phenomenological Fermi-liquid theory is well developed only for isotropic model of a metal. We start from the well known expressions for matrix elements of the Fermi-liquid kernel $ F_{\nu\nu'}^{\nu_1 \nu_2}$ \cite{17}:  
   \be 
 F_{\nu \nu'}^{\nu_1 \nu_2} =
\varphi_{\alpha \alpha'}^{\alpha_1 \alpha_2} 
\delta_{\sigma \sigma'} \delta_{\sigma_1 \sigma_2} +
4 \psi_{\alpha \alpha'}^{\alpha_1 \alpha_2} 
({\bf s}_{\sigma \sigma'} {\bf s}_{\sigma_1 \sigma_2}),
   \ee
 where $ \alpha $ is the set of the orbital quantum numbers, $ \sigma $ is the spin number, $ \bf s $ is the spin operator.

The FS under consideration is axialy symmetric. Therefore Fermi-liquid functions $ \varphi \bf (p, p') $ and $ \psi \bf (p, p') $ on the FS depend only on the cosine of the angle $ \theta $ between the vectors $ \bf p_\perp $ and $\bf p_\perp'$ and on the projections of the quasimomenta on the symmetry axis $ p_z, p_z´ .$ We can represent the function $ \varphi \bf (p,p') $ as a sum of even and odd contributions in $\cos  \theta: $
  \be 
\varphi {\bf (p,p')} = \varphi_0 (p_z,p_z',\cos\theta) + \cos \theta \varphi_1 (p_z,p_z',\cos\theta),
   \ee
  where $ \varphi_0 (p_z,p_z',\cos\theta) $ and $ \varphi_1 (p_z,p_z',\cos\theta) $ are even function in $ \cos \theta. $ By virtue of the FS invariance under replacements $ \bf p \to -p $ and $ \bf p'\to - p',$ the functions $ \varphi_0 (p_z,p_z',\cos\theta) $ and $ \varphi_1 (p_z,p_z',\cos\theta) $ should not vary under simultaneous change of sign of $ p_z $ and $ p_z'. $ It enables us to separate each of the mentioned functions into the parts which are even and odd in $ p_z p_z'$ and to rewrite Eq. (14) in the form:
      \be 
  \varphi {\bf (p,p')} = \varphi_{00} + p_z p_z'\varphi_{01} + \cos \theta (\varphi_{10} + p_z p_z' \varphi_{11} ).
   \ee
   Here, the functions $ \varphi_{00},\varphi_{01},\varphi_{10} $ and $ \varphi_{11} $ depend on $ p_z,p_z'$ and $ \cos \theta $ and are even functions in all their arguments. We can write a similar expression for $ \psi \bf (p,p'). $

To proceed we expand functions included into Eq. (15) in Fourier series in the angles $ \Phi $ and $ \Phi'.$ The functions are even in $ \cos \theta, $ therefore the expansions have the form:
   \be 
 \varphi_{ik} (p_z,p_z',\cos\theta) = \sum_{s=-\infty}^\infty
\varphi_{ik}^{|2s|} (p_z,p_z')e^{2is(\Phi-\Phi')}.
   \ee
  Here $i,k = 0,1. $ Coefficients $ \varphi_{ik}^{|2s|}(p_z,p_z'), $ in the expansions (16) are even functions in both arguments. Besides, they should not change when the arguments are interchanged. We can simplify further consideration assuming that:
   \be 
 \varphi_{ik}^{|2s|} (p_z,p_z') = \frac{1}{g_0} 
P_{ik}^{|2s|} (p_z) P_{ik}^{|2s|} (p_z'),
   \ee
  where both factors are even functions, and $ g_0 $ is the ``bare" DOS of the quasiparticles on the FS in the absence of the magnetic field $ \bf B. $
 
 The expansions of the functions $ \psi_{ik}^{|2s|} (p_z,p_z',\cos\theta) $ are similar to Eq. (16). The coefficients in these expansions also can be presented as:
 \be 
 \psi_{ik}^{|2s|} (p_z,p_z') = \frac{1}{g_0} 
Q_{ik}^{|2s|} (p_z) Q_{ik}^{|2s|} (p_z'),
   \ee

To calculate the quantity $ N_\zeta^* $ under the conditions of the semiclassical quantization $(\gamma \gg 1$ where $ \gamma^2 = 2\zeta/\hbar \Omega) $ we can replace the matrix elements $ \varphi_{\alpha\alpha'}^{\alpha_1\alpha_2} $ and $ \psi_{\alpha\alpha'}^{\alpha_1\alpha_2} $ by their semiclassical analogs. The analog for $ \varphi_{\alpha\alpha'}^{\alpha_1\alpha_2} $ is the Fourier component in the expansion of the function:
    \bea 
 \varphi {\bf (q;p,p')} &
=&\exp \left[- \frac{i c q_y p_\perp}{eB} \sin \Phi \right] \varphi {\bf (p,p')} \nn \\ &\times&
\exp \left[- \frac{i c q_y p_\perp'}{eB} \sin \Phi' \right],  
   \eea
  in the Fourier series in the angles $ \Phi $ and $ \Phi'.$
Here, $\varphi {\bf (p,p')} $ is the semiclassical Fermi-liquid function. With the help of the known identity for the Bessel functions:
   \be 
  \exp(\pm iz \sin \Phi) = \sum_{m=-\infty}^\infty J_m (z) \exp (\pm im \Phi),
   \ee
  we arrive at the result:
  \bea 
 \varphi_{ll'} {\bf (q;p,p')} &=& 
\frac{1}{4 \pi^2} \!\sum_{m=-\infty}^\infty \!
J_m \left(\frac{cq_yp_\perp}{eB} \right) 
 \nn \\ & \times& \!\!\!\!
 \sum_{m'=-\infty}^\infty \!\!J_{m'} \left(\frac{cq_yp_\perp'}{eB} \right)
 \!\int_0^{2\pi}\! d \Phi \exp[i(m-l] \Phi] 
 \nn \\ & \times& \!
\int_0^{2\pi} \!d \Phi' \exp[i(m'-l') \Phi'] \varphi \bf (p,p').
         \eea
  Substituting the expression (14) for $ \varphi \bf (p,p') $ into Eq. (21) and taking into account the expansions (16) we arrive at the final result:
  \bea 
  && \varphi_{ll'} ({\bf q;p,p'}) 
 \nn \\  &=&  \frac{1}{g_0} 
\sum \limits_{s = -\infty}^\infty 
\left \{ \left [ P_{00}^{|2 s|} (p_z) P_{00}^{|2 s|}(p'_z) 
 \right. \right.  \nn\\
   &+&  \left.
p_z p'_z P_{01}^{|2 s|} (p_z) P_{01}^{|2 s|}(p'_z) \right ] 
         \nn \\ 
&\times &  J_{l - 2s}
\left ( \frac{c q_y p_\perp}{e B} \right )
J_{l' - 2s} \left ( \frac{c q_y p_\perp' }{e B} \right ) 
     \nn \\ &+& 
\left [ P_{10}^{|2 s|} (p_z) P_{10}^{|2 s|}(p'_z) +
p_z p'_z P_{11}^{|2 s|} (p_z) P_{11}^{|2 s|}(p'_z) \right ] 
  \nn \\ &\times&  
 \left [ J_{l - 2s - 1}
\left ( \frac{c q_y p_\perp}{e B} \right )
J_{l' - 2s -1} \left ( \frac{c q_y p_\perp' }{e B} \right )
\right.                                                 
   \nn\\ &+& 
\left. \left.   J_{l - 2s + 1}
\left ( \frac{c q_y p_\perp}{e B} \right )
J_{l' - 2s +1} \left ( \frac{c q_y p_\perp' }{e B} \right )
\right ] \right \} .
   \eea
   In the expansions of $ \psi_{ll'}\bf (q;p,p') $ the functions $ P_{ik}^{|2s|} (p_z) $ and $ P_{ik}^{|2s|} (p_z') $ in (22)  have to be replaced by the functions $ Q_{ik}^{|2s|} (p_z) $ and $ Q_{ik}^{|2s|} (p_z') .$ The derived expressions (22) enable us to study any Fermi-liquid effects for conductors whose FSs are axially symmetric.

Here, we consider the most interesting case when the gradients of the deformation of the lattice are directed across $ {\bf b}_0 ({\bf q \perp b}_0) .$ Introducing the notations $ 2N = n + n'; \ 2N'= n_1 + n_2; \ l = n - n'; l'=n_1 -n_2 $ we can write the expansions for the matrix elements  $ \varphi ({\bf q}; N l p_z; \, N' l' p'_z) $ and $ \psi ({\bf q}; N l p_z; \, N' l' p'_z) .$ We arrive at these expansions  replacing the quantities $ p_\perp, p_\perp'$ in the arguments of Bessel functions in Eq. (22) by their quantum analogs $\hbar \sqrt{2N+1} $ and $\hbar \sqrt{2N'+1} $.

The functions $ P_{ik}^{|2 s|}(p_z) $ and $ Q_{ik}^{|2 s|}(p_z) $ are even, and the matrix elements $ n_{\nu \nu'}(\bf -q) $ are diagonal in spin number $ \sigma .$ Owing to this, one can take into account only the first term in the first square brackets in Eq. (22) in calculation of the quantity $ N_\zeta^*. $ The contributions from the other terms will be equal to zero. Thus in the following calculations we can assume that $
\varphi ({\bf q}; N l p_z; \, N' l' p'_z) $ and $ \psi ({\bf q}; N l p_z; \, N' l' p'_z) $ have the form
      \bea 
   \varphi ({\bf q}; N l p_z; \, N' l' p'_z)
& =&  \frac{1}{g_0} 
\sum \limits_{s = -\infty}^\infty 
 P_{00}^{|2 s|} (p_z) P_{00}^{|2 s|}(p'_z) 
             \nn \\
&\times &  J_{l - 2s}
\left ( \frac{\hbar c q }{e B}
\sqrt {2N + 1} \right ) 
  \nn\\ &\times& 
J_{l' - 2s} \left ( \frac{\hbar c q }{e B}
\sqrt {2N' + 1} \right ); 
   \nn\\
\psi ({\bf q}; N l p_z; \, N' l' p'_z)
& =&  \frac{1}{g_0} 
\sum \limits_{s = -\infty}^\infty 
 Q_{00}^{|2 s|} (p_z) Q_{00}^{|2 s|}(p'_z) 
            \nn \\
&\times &  J_{l - 2s}
\left ( \frac{\hbar c q }{e B}
\sqrt {2N + 1} \right ) \nn\\ &\times& 
J_{l' - 2s} \left ( \frac{\hbar c q }{e B}
\sqrt {2N' + 1} \right ).
   \eea
  Further we omit the lower indices of the functions $ P_{00}^{|2s|} (p_z) $ and $ Q_{00}^{|2s|} (p_z) $  for simplicity.

Under conditions of the semiclassical quantization, when $ \gamma \gg 1,$ we have for the arbitrary single particle operator $ \Phi_{\nu\nu'} $ (see Ref. \cite{13}):
  \bea 
&&\sum_{\nu\nu'} \frac{f_\nu - f_{\nu'}}{E_\nu - E_{\nu'}}
 \Phi_{\nu \nu'} 
\nn\\&=& -\frac{1}{4 \pi^2 \hbar^3}
\sum_l \sum_{\sigma\sigma'} \int dp_z \bigg [m_\perp(p_z) \Phi_l^{\sigma\sigma'} (p_z) 
  \nn \\   &+&  
 \delta_{l_0} \delta_{\sigma\sigma'} \sum_m (\Delta_m + \sigma \Delta_m') m_\perp (p_m) \Phi_l^{\sigma\sigma'} (p_m)\bigg].
     \eea
  Here, $ m_\perp (p_z) $ is the cyclotron mass which is assumed to be a constant in further calculations; $ p_m $ is the value of the longitudinal component of the quasimomentum, corresponding to the $m$th external cross section of the FS. The functions $ \Delta $ and $ \Delta'$ describe  oscillating corrections which appear in quantizing magnetic fields. Their form depends on the particular character of the energy-momentum relation for the quasiparticles. We will analyze these quantum oscillating corrections below within the framework of our model for the electron energy spectrum in organic metals. Integration with respect to $ p_z $ in Eq. (24) is carried out within the limits determined by the shape and size of the FS.

Using the asymptotic formula (24), and the expansions (23) we can obtain the expression for the renormalized DOS: 
   \be 
N_\zeta^* = g_0
\left [ 1 - \alpha_0 + 
\frac{(1 - \overline \alpha_0)^2[\Delta + \beta (\Delta^2 - \Delta'^2)]}{1 + (\alpha + \beta)\Delta +
\alpha \beta (\Delta^2 - \Delta'^2)} \right ] .
    \ee 
  Here:
 \bea 
\alpha_0 & =& \frac{A_0}{1 + \tilde A_0} ;
\qquad
\overline \alpha_0  = \frac{\overline A_0}
{1 + \tilde A_0} ;
    \nn\\
\alpha &=& \frac{A_0^*}{1 + \tilde A_0} ;
\qquad
\beta  = \frac{B_0^*}{1 + 
\tilde B_0} ;                                               
 \nn\\
A_0 &=& \frac{m_\perp}{2 \pi^2 \hbar^3 g_0}
\left ( \int P^0 (p_z) d p_z \right )^2;
 \nn\\
\overline A_0 &=& \frac{P^0 (0) m_\perp}{2 \pi^2 \hbar^3 g_0} 
\int  P^0 (p_z) d p_z , 
                                      \nn\\
 \tilde A_0 &=& \frac{m_\perp}{2 \pi^2 \hbar^3 g_0}
\int  [ P^0 (p_z)]^2 d p_z .
   \nn\\
 \tilde B_0 &=& \frac{m_\perp}{2 \pi^2 \hbar^3 g_0}
\int  [ Q^0 (p_z)]^2 d p_z ;
\nn\\
 A_0^* &=& [P^0 (0)]^2;
 \qquad  B_0^* =[Q^0 (0)]^2.
     \eea
Neglecting  terms proportional to the differencies $ \Delta^2 -
\Delta'^2 $ in the numerator and the denominator of the Eq. (26) we can simplify the expression for the renormalized density of states: 
    \be 
N_\zeta^* = g_0
\left [ 1 - \alpha_0 + \frac{(1 - \overline \alpha_0)^2 \Delta} {1 + (\alpha + \beta)\Delta } \right ] .
        \ee
 The expression (27) is easily generalized  to cover the case when the Fermi surface has several extremal cross sections. In this case we obtain
                   \be 
N_\zeta^* = g_0\!\!
\left [ \!1 - \alpha_0 +  \frac
{\sum \limits_m (1 - \overline \alpha_m)^2 \Delta_m}
{1 + \sum \limits_m (\alpha_m + \beta_m) \Delta_m } \right ]\!\!
\equiv g_0(1-\alpha_0 +K) .
        \ee
 Here, summation over $ m $ is carried out over the extremal cross sections:
  \bea 
  &&
\alpha_m = \frac{A_{m}^*}{1 + \tilde A_0} \, ;
\qquad
\overline \alpha_m = \frac{\overline A_m}{1 + \tilde A_0} \, ;
\qquad
\beta_m = \frac{B_{m}^*}{1 + \tilde B_0} \, ;
   \nn\\&&
A_{m}^* = [P^0 (p_m)]^2;
\qquad
B_{ m}^* = [Q^0 (p_m)]^2;
 \nn\\ &&
\overline A_m =  \frac{m_\perp}{2 \pi^2 \hbar^3 g_0} \, P^0 (p_m) \int P^0 (p_z) d p_z .
        \eea
  Neglecting all terms arising due to the interaction among quasiparticles, we arrive at the result for the ``bare" DOS:
  \be 
 N_\zeta = g_0 (1+ \Delta).
   \ee

The oscillating term $ \Delta $ describes  quantum oscillations of the ``bare" DOS. This term is small compared to unity when the temperature is moderately low $ (\theta \sim 1). $ When $ \theta \ll 1 $ the magnitude of the function $ \Delta $ can reach values of the order of unity. Comparison of Eqs. (28) and (30) shows that there are pronounced distinctions between the ``bare" DOS $ N_\zeta $ and the DOS $ N_\zeta^* $ renormalized due to the interaction among quasiparticles within low temperature range. Under $ \theta \ll 1 $ the amplitude and the form of the function $ K $ which describes the quantum oscillations of $ N_\zeta^* $ can differ significantly from the corresponding characteristics of the function $ \Delta. $ In particular, the denominator in Eq. (22) can turn zero at the peaks of the oscillations. Thus, the function $ K $ has singularities caused by the Fermi-liquid interaction.

The specific features of the low temperature quantum oscillations in renormalized density of states have to be manifested in oscillations of the thermodynamic observables. We consider, for example, the oscillating contribution to the magnetic susceptibility for a uniform field $ \bf B.$ To simplify the following calculations we assume that the unique extreme cross-section of the axisymmetric FS is located at $ p_z = 0. $ We remark here that possible singularities in the longitudinal magnetic susceptibility near the peaks of quantum oscillations of the renormalized DOS were first analyzed in Ref. \cite{18} for an isotropic electron liquid (see also Ref. \cite{19}). Generalizing the corresponding result of Ref. \cite{18}, we can show  that in the case of the axisymmetric FS, $ \chi_{||} $ is proportional to the oscillating function $ K': $
    \bea 
  K'&=& \frac{K}{1 + [1 - \overline \alpha_0 (0)] K}
   \nn\\ &=&
\frac{\Delta (0)}{1 + [1 + \alpha_0 (0) - \overline \alpha_0 (0) + \beta_0 (0)] \Delta (0)} .
   \eea   
 
Within the low temperature range the amplitude of the oscillations in $ \Delta (0) $ might increase to such an extend that it reaches the values of the order of unity. Under these conditions the form of the de Haas-van Alphen oscillations is determined by the denominator in Eq. (31) to a great extent, and it depends critically on the Fermi-liquid interaction. When $[1 + \alpha_0 (0) - \overline \alpha_0 (0) + \beta_0 (0)] < 0 $ the denominator in Eq. (31) can go to zero at the peaks of the oscillations and the susceptibility $ \chi_{||} $ would increase beyond all bounds. It would lead to the magnetic instability $ (\chi_{||} < 1 /4\pi) $ at the peaks of the oscillations (see Fig. 1).  Thus, the low temperature analog for the Shoenberg effect can be observed. It follows from the results of Ref. \cite{18} as well as from our expression (31) that the magnetic susceptibility $ \chi_{||} $ of noninteracting 3D electron gas can exhibit a series of maxima of a finite magnitude when temperature goes to zero. G. Montambaux {\it et al.} \cite{11} came to a similar conclusion irrespective of the earlier results of Ref. \cite{18}.

\begin{figure}[t]  
\begin{center}
\includegraphics[width=3.3cm,height=4.5cm]{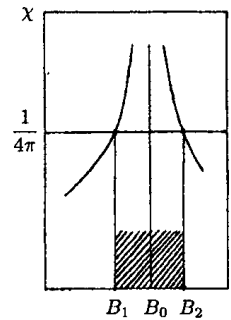}
\caption{ Dependence of the magnetic susceptibility $ \chi_{||} $ on the magnetic field at $ T = 0. $ The domain of the values of the magnetic field where $ 4 \pi \chi > 1 $ is section-lined.
}  
\label{rateI}
\end{center}
\end{figure}

These singularities of the renormalized DOS $ N_\zeta^* $ originating due to the Fermi-liquid interaction can give rise to specific features in the quantum oscillations of thermodynamic observables in layered conductors. However, their manifestations would depend on the local geometry of the FS near its extreme cross sections. To analyze these anomalies in more detail we introduce a concrete model of the FS for layered organic metals which allows us to consider a broad class of the FSs of various profiles.

\section{4. model and results}

The usual approach in theoretical studies of electron properties of layered conductors is to keep only the first few terms in the sum over $ k $ in Eq. (2). As a rule (see Refs. \cite{20,21,22}, only the first term is taken into account. This corresponds to results obtained in the tight-binding approximation. Here, we use a different approach to describe the electron energy spectrum of the charge carriers in layered conductors, whose FS is defined by the equation
   \be 
 E {\bf (p)} = \frac{p_\perp^2}{2 m_\perp} - \eta v_0 p_0 E \left (\frac{p_z}{p_0} \right),
   \ee
  where $ v_0 = (2 \zeta /m_\perp )^{1/2},\ E (p_z/p_0) $ is an even function periodic in its argument $ p_z/p_0 $ with a period equal to 2, and $ \eta $ is a dimensionless parameter characterizing the rate of rippling of the FS. The quantity $ - \eta v_0 p_0 E (p_z/P_0) $ is the sum of the trigonometric series in (2). By selecting the type of this function we can obtain FS's shaped as pinched cylinders with different profiles. This approach provides broad possibilities in analyzing the effect of the shape of the Fermi surface on observed characteristics of layered conductors.

Let us assume that the function $ E (p_z/p_0 )$ in the interval $ - p_0 \leq p_z \leq p_0 $ is described by the expression
   \be 
 E \left (\frac{p_z}{p_0}\right) = \frac{1}{r l} \left[1-\left|\frac{p_z}{p_0}\right|^l \right]^r,
   \ee
  where the parameters $l $ and $ r $ take values more than unity. The model specified by (32) and (33) makes it possible to describe a broad class of FSs.

The Gaussian curvature of the FS is described by the expression
  \be 
  \kappa (p_z) = \frac{m_\perp^2[v_z^2 + (p_\perp^2/m_\perp) \partial v_z/\partial p_z]}{(p_\perp^2 + m_\perp^2 v_z^2)^2} .
 \ee
  At $ l = r=2 $ the curvature of the FS at its sections by the planes $ p_z = 0 $ and $ p_z = \pm p_0 $ equals:
   \bea 
  \kappa(0) &=& \frac{\delta S}{S_{\max}} \frac{1}{p_0^2};
 \\
 \kappa(\pm p_0) &=& - \frac{2 \delta S}{S_{\min}} \frac{1}{p_0^2},
  \eea
 where $ S_{\max} $ and $ S_{\min} $ are the maximum and minimum sectional areas of the FS: $ S_{\max} = S(0),\ S_{\min} = S(\pm p_0), $ and $ \delta S = S_{\max} - S_{\min} = (\pi/2) m_\perp \eta v_0 p_0. $ Thus, if the FS remains a pinched cylinder $(\eta \neq 0), $ its curvature at all points of the sections with extremal diameters is finite. Similar results are obtained if the tight-binding approximation is used to describe the electron energy spectrum.

For $ r \neq 2$ and $ l = 2 $ the curvature of the FS near $ p_z \pm p_0 $ remains finite and $ \kappa (0) $ is still described by (35). However, the asymptotic behavior of the curvature of the FS near $ p_z = \pm p_0 $ is different: 
  \be 
  \kappa (p_z) = - 2 (r-1) \frac{\delta S}{S_{\min}} 
\frac{1}{p_0^2} \left[1-\left(\frac{p_z}{p_0}\right)^2 \right]^{r-2} .
   \ee
 Thus, for $ 1<r<2 $ the curvature of the FS has singularities in these sections. For $ p_z = \pm p_0 $ the curvature $ \kappa (p_z )$ vanishes at $ p_z = \pm p_0. $ The corresponding sections of the Fermi surface are lines of parabolic points. The larger the value of the parameter $r,$ the more the FS near these sections resembles a cylinder.

\begin{figure}[t]  
\begin{center}
\includegraphics[width=4cm,height=7cm]{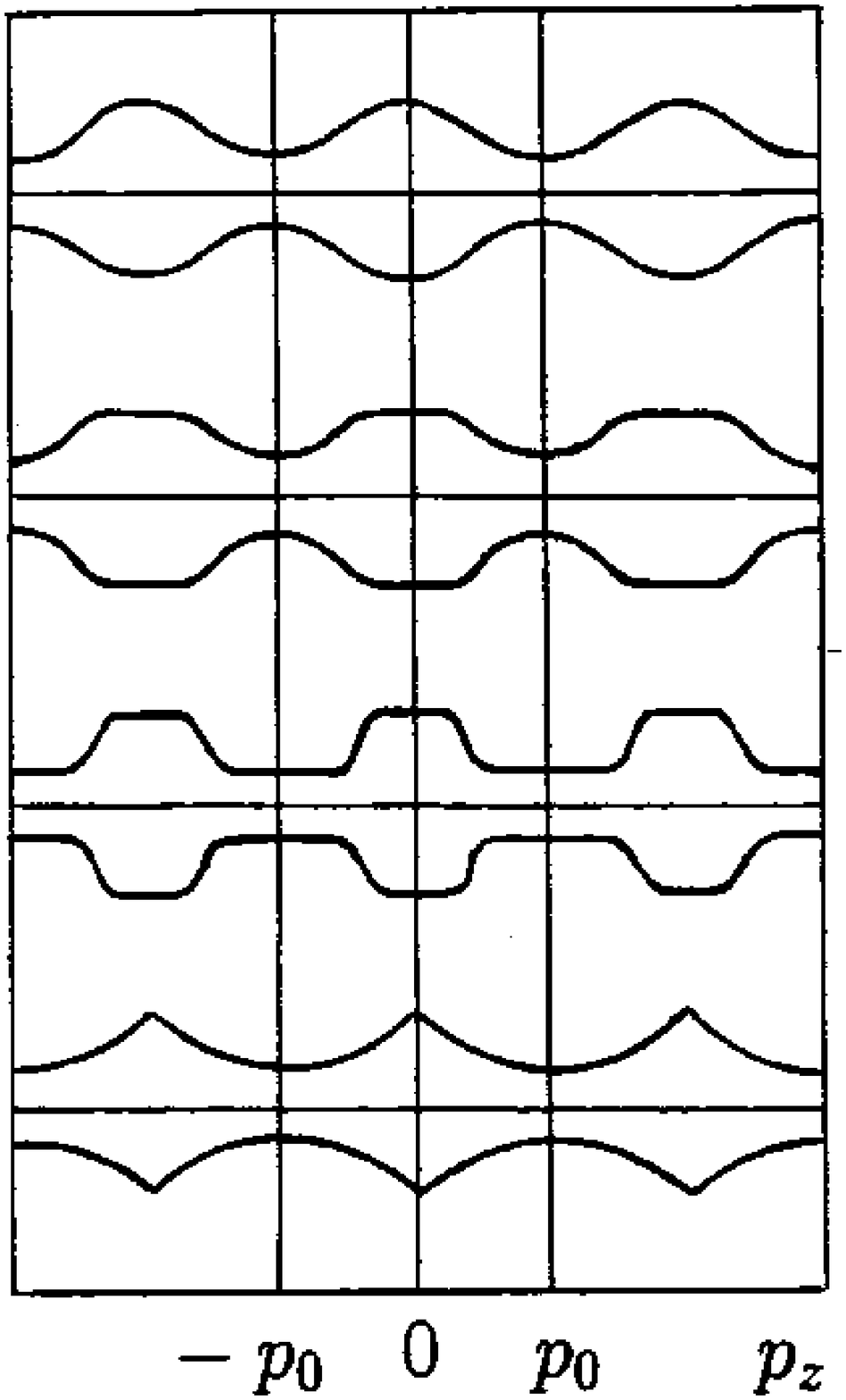}
\caption{ Profiles of corrugated cylinders described by Eqs.(32) and (33), for different values of the parameters $ r $ and $ l; \, \mbox{(a)}\ l=r=2; \, \mbox{(b)} \, r=2,l>2;\, \mbox{(c)}\, r>2,l>2; \, \mbox{(d)}\, r=2, 1<l<2. $ 
}  
\label{rateI}
\end{center}
\end{figure}

{\small
\begin{picture}(0,0)(0,180) 
\put(62,484){(a)}
\put(62,444){(b)}
\put(62,401){(c)}
\put(62,359){(d)}
\end{picture}}
\vspace{0mm}

The anomalies in the curvature of the FS near $ p_z = 0 $ can be described by the model (32) and (33) with $ r = 2 $ and $ l \neq 2. $ Here the curvature of the FS near $ p_z = 0 $ is described by the asymptotic expression
  \be 
  \kappa(p_z) = (l-1) \frac{\delta S}{S_{\max}}\frac{1}{p_0^2} \left|\frac{p_z}{p_0} \right |^{l-2}.
  \ee 

For $ 1 <l<2 $ the curvature of the FS has a singularity at $ p_z =0; $ if $ l > 2 $ the Fermi surface near $ p_z = 0 $ resembles a cylinder, and the larger the value of $ l $ the
closer the resemblance. Finally, if $ r \neq 2$ and  $ l \neq2, $ we have a surface in the form of a pinched cylinder with curvature singularities in all the sections with extremal diameters. The profiles of the Fermi surface described by (32) and (33) are depicted schematically in Fig. 2.

We shall use our model to obtain the expressions for the oscillating function $ \Delta. $ The form of the function $ \Delta $ depends importanty on the characteristic properties of the electron energy spectrum. Assuming that FS curvature on the effective strips is nonzero $(r=l=2) $ we can obtain:
   \bea 
\Delta  &=& \Delta(0) + \Delta(p_0) 
=\frac{1}{\gamma} \sqrt{\frac{S_0}{\delta S}} \left \{
\sum_{n=1}^\infty \frac{(-1)^n}{\sqrt n}\psi_n (\theta) \right.
  \nn\\ &\times&
\cos \left(\frac{nc S_{\max}}{\hbar e B} - \frac{\pi}{4} \right ) \cos\left(\pi n \frac{\Omega_0}{\Omega} \right) +
 \sum_{n=1}^\infty \frac{(-1)^n}{\sqrt n}\psi_n (\theta)
  \nn\\&\times& \left.
\cos \left(\frac{nc S_{\min}}{\hbar e B} +\frac{\pi}{4} \right ) \cos\left(\pi n \frac{\Omega_0}{\Omega} \right) \right\}.
   \eea
   Here, $S_0 = 2 \pi m_\perp \zeta $ is the area of the FS cross-section in the limit $ \eta \to 0, $ when the FS turns a pure cylinder; $ \hbar \Omega_0 $ is the spin splitting energy.

In the case when there is an anomaly in the FS curvature at $ p_z = \pm p_0 \ (l = 2; r\neq 2) $ the contribution to the oscillating function $ \Delta $ from the vicinities of corresponding cross-sections assumes a form:
   \bea 
  \Delta(p_0) &=& \left(\frac{rS_0}{2 \gamma^2 \delta S} \right)^{1/r} \frac{\Gamma (1/r)}{r} \sum_{n=1}^\infty \frac{(-1)^n}{n^{1/r}} \psi_n(\theta)
 \nn\\&\times& 
\cos \left(\frac{nc S_{\min}}{\hbar e B} +\frac{\pi}{2r} \right ) \cos\left(\pi n \frac{\Omega_0}{\Omega} \right),
     \eea
 where $ \Gamma $ is the gamma function. Under this condition the first term in Eq. (40) retains its form. Correspondingly when the local geometry of the FS is characterized with the anomalous curvature at $ p_z = 0 \ (r = 2; l\neq2) $ the oscillating function $ \Delta (0) $ is described by the expression:
   \bea 
  \Delta(0) &=&  \left (\frac{l S_0}{2 \gamma^2 \delta S}\right)^{1/l} \frac{\Gamma (1/l)}{l} \sum_{n=1}^\infty \frac{(-1)^n}{n^{1/l}} \psi_n(\theta)
 \nn\\ &\times& 
\cos \left(\frac{nc S_{\max}}{\hbar|e| B} -\frac{\pi}{2l} \right ) \cos\left(\pi n \frac{\Omega_0}{\Omega} \right),
   \eea   
  At the same time the expression (40) for $ \Delta (p_0) $ remains valid.

When the effective strip on the FS centered at $ p_z = 0 $ or $ p_z = \pm p_0$ is close to a cylinder in shape, its contribution to the function $ \Delta $ has the same form as for the conductor with two-dimensional energy spectrum. For instance, if $ l \gg 1$ the expression (41) can be transformed to the form
   \be 
  \Delta (0) = \sum_{n=1}^\infty (-1)^n \psi_n(\theta)
\cos \left(\frac{nc S_{\max}}{\hbar |e|B} \right)\cos \left(\pi n \frac{\Omega_0}{\Omega} \right).
   \ee
  In the case when both parameters $(r$ and $l)$ are large compared to unity the expression for $ \Delta $ contains two terms of the form (42). Both terms represent the contributions from the quasicylindrical strips on the FS. These oscillating terms have to differ in period. The difference in their periods arises due to the distinction in the extremal cross sectional areas. It may be noticeable if the magnetic field is not too strong and the inequality $ \gamma^2 \delta S/S_0 \gg 1 $ is satisfied.

In a very strong magnetic field the inequality $ \gamma^2\delta S /S_0 \ll 1 $ is satisfied. Under this condition the difference in form between the FS and the cylinder does not affect the oscillating function $ \Delta: $
   \be 
 \Delta = 2 \sum_{n=1}^\infty (-1)^n \psi_n (\theta) \cos(\pi n \gamma^2) \cos \left(\pi n \frac{\Omega_0}{\Omega} \right).
   \ee
 
When a temperature is moderately low $(\theta \sim1) $ we can keep only the first term in the sum over $``n"$ in Eqs. (39)--(43). Thus, under $ \theta \sim 1, $ the dependence of the function $ \Delta $ on the inverse magnetic field has to be of a harmonic type. Because of the factor $ \exp (-\theta) ,$ the amplitude of the oscillations of the function $ \Delta $ under $ \theta \sim 1 $ is small compared to unity even in strong magnetic fields.

When $ \theta \ll 1, $ the form of the oscillations becomes much more complicated and their amplitude increases. To evaluate the amplitude of the oscillations under this condition one can use the asymptotic formulae following from the Euler-Maclaurin summation formula. As a result, under this condition we obtain the following estimation for the amplitude of the oscillations:
   \bea 
 \Delta_{\max} (p_0) &\approx& \left (\frac{S_0 \theta}{\gamma^2 \delta S} \right)^{1/r} \frac{1}{\theta},  
    \\
\Delta_{\max} (0) &\approx& \left (\frac{S_0 \theta}{\gamma^2 \delta S} \right)^{1/l} \frac{1}{\theta}.
  \eea
  Here, $ r $ and $ l $ are the dimensionless parameters whose values determine the local geometry of the FS in our model (33). In the case when the FS curvature remains finite and nonzero at $ p_z = 0 $ and $ p_z = \pm p_0 $ we have:
 \bea 
  \Delta_{\max} (p_0) =\Delta_{\max} (0) &=& \frac{1}{\gamma}
 \sqrt{\frac{S_0}{\delta S}} \sum_{n=1}^\infty \frac{1}{2\sqrt{2n}} \psi_n(\theta) \nn\\
 & \approx &
 \frac{3}{2}\frac{1}{\gamma \sqrt \theta}  \sqrt{\frac{S_0}{\delta S}}.
  \eea
  For the same values of the parameters $ \gamma\, , \delta S/S_0 $ and $ \theta $ the ratio of the amplitudes of oscillations described with the formulae (41) and (39) is of the order of $ (\gamma^2\delta S/S_0 \theta)^{(l-2)/2l}. $ Thus, under typical experimental conditions when $ \gamma^2 \delta S/S_0 \gg1, $ the oscillations associated with a quasicylindrical extremal cross-section of the FS have an amplitude much larger than the amplitude of the ordinary oscillations described by formula (39). If $ 1<l <2 $ one of the principal radii of curvature vanishes at all points of the extremal cross-section of the Fermi surface at $ p_z = 0. $ In this case the amplitude of oscillations described by formula (39) is much smaller than the amplitude of the oscillating parts of the contributions from the FS ordinary extremal cross-sections.

When $ \theta \ll 1$ the oscillation amplitude  may be of the order of unity. Charge carriers concentration in layered organic metals is far below than in ordinary metals. This generates much more favorable conditions for observation of the salient features in low temperature quantum oscillations in thermodynamic quantities. If we take for $ m_\perp $ and $S_0 $ the values obtained in experimental study of the Fermi surface of the organic metal $\beta-\mbox{(ET)}_2\mbox{IBr}_2 $ (Ref. \cite{23}) $(m_\perp \approx 4.5 m_0;\ S_0 \approx 1.26\times 10^{-39}$ g$^2$cm$^2/$s$^2;\ m_0$ is free electron mass) the condition $(3/2) 1/\gamma\sqrt\theta\sqrt{S_0/\delta S} \sim 1$ in magnetic fields $ B\sim 200$ kG and for $ \delta S/S_0 \approx 0.05 $ will be satisfied at temperatures of the order of one Kelvin.

Now we can return back to our expressions (11) and (12) for the elastic moduli and to proceed the analysis of their low temperature anomalies in layered conductors. For definiteness, we assume that the FS of the considered organic metal is nearly cylindrical in shape near $ p_z = 0 $ and its curvature at $ p_z =\pm p_0 $ is finite and nonzero. Under these assumptions the oscillating correction $ \Delta (0) $ predominates and we can write:
  \be 
 \lambda_0 =\frac{N^2}{N_\zeta^*} = \frac{N^2 [1-(1- \overline \alpha_0 (0)) K']}{(1 - \alpha_0)g_0},
        \ee 
   \be
 \lambda = \lambda_0 \left (1 + \frac{4 \pi \chi_\zeta}{1 - 4 \pi \chi_{||}} \right) =
 \frac{N^2 [1-(1- \overline \alpha_0 (0)) K'']}{ (1 - \alpha_0) g_0}.
  \ee
  Here, the function $ K''$ includes the oscillating term $ \Delta (0), $ therefore it oscillates itself in a strong magnetic field:
  \be 
  K'' = \frac{\Delta (0)}{1 +[1 + \alpha_0 (0) - \overline \alpha_0 (0) + \beta_0 (0) - 4 \pi \chi_0 \gamma^4] \Delta (0)}
 .  \ee
  To obtain the expression for $K''$ we used  the result (31) for the longitudinal part of the magnetic susceptibility $ 
\chi_{||}; $ the quantity $ \chi_0 $ is related to the Landau diamagnetic susceptibility. The latter equals $ - \chi_0/3. $

The sharpest difference between $ \lambda $ and $ \lambda_0 $ is displayed at those values of the magnetic field $ \bf B, $ at which the quantity $ 1 - 4 \pi \chi_{||} $ goes to zero and the denominator in Eq. (49) turns  zero at the peaks of the quantum oscillations. It makes the elastic constant $ \lambda $ (or the electron contribution to the elastic modulus $c_{11})$ vanish near the oscillations maxima. Thus, there can arise lattice instability. It is connected with the instability in the magnetization and arises because of the magnetostriction. Under these conditions the Fermi-liquid correlation plays the major part. When $ 4 \pi \chi_0 \gamma^4 < 1 $ the instability can arise only under $ \alpha_0 (0) - \overline \alpha_0 (0) + \beta_0 (0) < 0. $ 

The lattice instability can, in principle, be displayed in the case, when the magnetostriction does not play a part (the strain gradient is directed along $\bf B),$ as the function $K$ [see Eq. (27)] in the low temperature region can also turn to infinity. Such instability was predicted in time in Refs. \cite{23} and \cite{24}. However, actually it is probably unattainable. The reason is that when $ \Delta $ increases the 
quantity $ 1 + [1 + \alpha_0 (0) - \overline \alpha_0 (0) + \beta_0 (0) - 4 \pi \kappa \gamma^4] \Delta (0) $ reaches  zero before the quantity $ 1 + [1 + \alpha_0 (0) - \overline \alpha_0 (0) + \beta_0 (0)] \Delta (0). $ Hence, the jump of the magnetic state will take place before the constant $ \lambda_0 $ runs into zero.

\section{5. Concluding Remarks}

The point of the present work is that the Fermi-liquid correlation among the quasiparticles can strongly affect their renormalized DOS. This was analyzed in detail for isotropic electron liquid \cite{13,14,15,18}. Here, we generalize this analysis to cover a broad class of layered conductors whose FSs are supposed to be axially symmetric. It is shown as a result of lengthy but straightforward calculations that $ N_\zeta^* $ significantly differs from the ``bare"DOS $N_\zeta $ under the condition that oscillating corrections $ \Delta_m $ which arise in the presence of the quantizing magnetic field are not small in magnitude compared to unity. Under this condition the oscillating part of $ N_\zeta^* $ can have singularities in the peaks of quantum oscillations. These singularities occur due to the Fermi-liquid interaction. The singularities of $ N_\zeta^* $ can cause specific features manifested in the oscillations of observables. In particular, they can lead to the singularities of the longitudinal magnetic susceptibility $ \chi_{||}$ at the magnetic fields corresponding to the peaks of the quantum oscillations of DOS, which leads to the violation of the stability of the considered substance $(\chi_{||}<1/4 \pi).$ A similar effect was briefly discussed in Ref. \cite{11}. Besides we obtain a possible singularity in the electron contribution to the elastic constant $ c_{22}$ which is connected with the instability in the magnetization and can give rise to the lattice instability of a special kind.

These interesting phenomena can be available for observation in organic metals (including substances of the $\alpha-\mbox{(BEDT-TTF)}_2\mbox{MHg(SCH)}_4 $ group) because of the specific character of the energy spectra of the charge carriers in these substances. It is too early to draw any conclusions about the local features of the geometry of the FS of the majority of layered organic metals, since there is a lot to study in the electron energy spectra of such materials. It can be assumed, however, that here, as in ordinary metals, the FS contains quasicylindrical bands or sections with an anomalously large curvature. These features of the local geometry of the FS can be created (if they are absent) or enhanced by applying an agent that changes the shape of the constant-energy surfaces, e.g. by applying external pressure along the normal to the conducting planes.

The above analysis shows that the special features in the profile of the rippled cylinder, which is the main part of the FS of layered organic metals, can substantially influence quantum oscillations of thermodynamic observables in these materials. The model developed in this paper makes it possible to study in detail the observable manifestations of the local geometry of the FS of layered conductors. It resolves some of difficulties that emerge when we use the model of tightly bound electrons. For instance, the characteristic features of the observable properties of layered conductors, arising due to strong anisotropy of the electrical conductivity can be described and analyzed without passing to the limit $ \eta \to 0, $ which corresponds to a conductor with a two-dimensional spectrum of the charge carriers.

The model specified by (32) and (33) enables us to carry out a detailed analysis of quantum oscillations of the elastic constants in organic metals and this leads us to the conclusion that the local geometry of their Fermi surfaces can provide favorable conditions for observation of the above described magnetic and lattice instabilities.

\section{Acknowledgments}

We thank G.M. Zimbovsky for help with the manuscript.

\section{ Appendix }
 
Here we present in more detail calculation of the renormalized DOS $ N_\zeta^* .$ We start from the expansion (23) for the matrix elements of the Fermi-liquid kernel. To proceed we use the notations:
       \bea 
I_s &=&
\sum \limits_{\nu \nu'}
\frac{f_\nu - f_{\nu'}}{E_\nu - E_{\nu'}} \, P^{|2 s|} (p_z)
J_{l - 2s} \left ( \frac{\hbar c q}{e B} \, \sqrt {2N + 1} \right ) \nn\\ &\times&
n_{\nu \nu'}^* (-q) \delta_{\sigma \sigma'} \delta_{x_0; x'_0 -
\hbar c q/e B} ,
        \\\nn\\
I'_s &=&  \sum \limits_{\nu \nu'}
\frac{f_\nu - f_{\nu'}}{E_\nu - E_{\nu'}} \, Q^{|2 s|} (p_z)
J_{l - 2s} \left ( \frac{\hbar c q}{e B} \, \sqrt {2N + 1} \right ) \nn\\ &\times&
n_{\nu \nu'}^* (-q) \delta_{\sigma \sigma'} \delta_{x_0; x'_0 -
\hbar c q/e B} ,
   \\\nn\\                      
g_{ss'} &=& - \frac{1}{g_0} \sum \limits_{\nu \nu'}
\frac{f_\nu - f_{\nu'}}{E_\nu - E_{\nu'}} \,
P^{|2 s|} (p_z) P^{|2 s'|} (p_z) 
     \nn\\                          & \times &
J_{l - 2s} \left ( \frac{\hbar c q}{e B} \, \sqrt {2N + 1} \right ) J_{l - 2s'} \left ( \frac{\hbar c q}{e B} \, 
\sqrt {2N + 1} \right )
 \nn \\ & \times & 
\delta_{\sigma \sigma'}  \delta_{x_0; x'_0 - \hbar c q/e B} ,
                    \\\nn\\   
q_{ss'}& =&  - \frac{1}{g_0} \sum \limits_{\nu \nu'}
\frac{f_\nu - f_{\nu'}}{E_\nu - E_{\nu'}} \,
Q^{|2 s|} (p_z) Q^{|2 s'|} (p_z) \nn\\
    \nn\\            &  \times&
J_{l - 2s} \left ( \frac{\hbar c q}{e B} \, \sqrt {2N + 1} \right )J_{l - 2s'} \left ( \frac{\hbar c q}{e B} \, \sqrt {2N + 1} \right )  \nn\\ &\times&
\delta_{\sigma \sigma'} \delta_{x_0; x'_0 -
\hbar c q/eB} ,
                    \\\nn\\
r_{ss'} &=&  - \frac{1}{g_0} \sum \limits_{\nu \nu'}
\frac{f_\nu - f_{\nu'}}{E_\nu - E_{\nu'}} \,
P^{|2 s|} (p_z) Q^{|2 s'|} (p_z) \times
    \nn\\ &     \times&
J_{l - 2s} \left ( \frac{\hbar c q}{e B} \, \sqrt {2N + 1} \right ) J_{l - 2s'} \left ( \frac{\hbar c q}{e B} \, \sqrt {2N + 1} \right )   \nn\\ &\times&
\delta_{\sigma \sigma'} \delta_{x_0; x'_0 -\hbar c q/eB} ,
     \\\nn\\
N_s &=& \sum \limits_{\nu \nu'}
\frac{f_\nu - f_{\nu'}}{E_\nu - E_{\nu'}} \, P^{|2 s|} (p_z)
   \nn\\ &\times&
J_{l - 2s} \left ( \frac{\hbar c q}{e B} \, \sqrt {2N + 1} \right )
n_{\nu' \nu} \bf (q) ;
    \\\nn\\    
N'_s &=& \sum \limits_{\nu \nu'}
\frac{f_\nu - f_{\nu'}}{E_\nu - E_{\nu'}} \, P^{|2 s|} (p_z)
  \nn\\ &\times&
J_{l - 2s} \left ( \frac{\hbar c q}{e B} \, \sqrt {2N + 1} \right )
\sigma n_{\nu' \nu} \bf (q) .
                                                 \eea

The expressions for the renormalized matrix elements
$ n_{\nu \nu'}^* \bf(-q) $ have the form
   $$
n_{\nu \nu'}^* (-{\bf q}) = n_{N l p_z}^* {\bf (-q)}
\delta_{\sigma \sigma'} \delta_{x_0; x'_0 - \hbar c q/e B}
\delta_{p_zp'_z},
                                                 $$
 where the quantity $ n_{N l p_z}^* (-q) $ satisfies the relation, following from the definition (7):
    \bea 
&& n_{N l p_z}^* {\bf (-q)} = n_{N l p_z} ({\bf -q}) 
\nn\\ &+&
\frac{1}{g_0} \sum_s P^{|2 s|} (p_z) 
 J_{l - 2s} \left ( \frac{\hbar c q}{e B} \, \sqrt {2N + 1} \right ) I_s  \nn\\ &+& 
\frac{\sigma}{g_0} \sum_s Q^{|2 s|} (p_z) 
J_{l - 2s} \left ( \frac{\hbar c q}{e B} \, \sqrt {2N + 1} \right ) I'_s \, .
                                                 \eea

When the Fermi surface is axisymmetric the matrix elements
$ n_{N l p_z}^* (-q)  $ do not depend on $ p_z :$
  \be 
n_{N l p_z} {\bf (-q)} =  J_{l}
\left ( \frac{\hbar c q}{e B} \, \sqrt {2N + 1} \right )  .
                                                 \ee 
 The average quantities $ I_s $ and $ I'_s $ satisfy a system of linear equations which follows from the Eq. (57):
  \be 
\left \{
\begin {array}    {l}
I_s + \sum \limits_{s'}(g_{ss'} I_{s'} + r_{ss'} I'_{s'} ) = N_s , \\                                                      
I'_s + \sum \limits_{s'}(r_{s's} I_{s'} + q_{ss'} I'_{s'} ) = N'_s .
\end{array}          \right.
                                         \ee
   Using (24) we get the expressions for the coefficients of the system (59):
    \bea 
g_{ss'} &=& \tilde A_{2s} \delta_{ss'}  +
\sum \limits_m \Delta_m P^{|2s|}(p_m) P^{|2s'|} (p_m)
\nn\\ &\times& J_{-2s}(qR_m) J_{-2s'}(qR_m) ,
          \\\nn\\                                  
q_{ss'}& =& \tilde B_{2s} \delta_{ss'}  +
\sum \limits_m \Delta_m Q^{|2s|} (p_m) Q^{|2s'|} (p_m)
\nn \\&\times& J_{-2s}(qR_m) J_{-2s'}(qR_m) ,
                \\\nn\\                   
r_{ss'}& =& \sum \limits_m \Delta'_m
P^{|2s|} (p_m) Q^{|2s'|} (p_m)
\nn \\&\times& J_{-2s}(qR_m) J_{-2s'}(qR_m) ,
                     \eea
where $ R_m $ in the radius of the cyclotron orbit corresponding to the
$ m $-th extremal cross section of the FS:
  \bea 
\tilde A_{2s}& =& \frac{m_\perp}{2 \pi^2 \hbar^3} \, \frac{1}{g_0} \int [P^{|2s|} (p_z)]^2 d p_z ; 
      \nn\\
\tilde B_{2s} & =& \frac{m_\perp}{2 \pi^2 \hbar^3} \, \frac{1}{g_0}
\int [Q^{|2s|} (p_z)]^2 d p_z .
        \eea
 In obtaining Eqs. (60)-(62) we used the identity concerning the
Bessel functions
   \be 
\sum \limits_{l = -\infty}^\infty J_{l - m}(x) J_l(x)
= \delta_{l 0} .
        \ee

 When the FS has the unique extremal cross-section
(at $ p_z = 0) $ Eq. (64) takes the form:
                       
  \be 
\left \{
\begin {array}    {lll}
\ds \frac{N_s}{1 + \tilde A_{2s}} &=
 I_s & +\ \ds \Delta \frac{P^{|2s|} (0) J_{-2s}(qR_{ex})}{1 + \tilde A_{2s}} X \nn \\ && +\ \ds
\Delta' \frac{P^{|2s|} (0) J_{-2s}(qR_{ex})}{1 + \tilde A_{2s}}  Y ,
  \nn     \\ \nn\\
 \ds \frac{N'_s}{1 + \tilde B_{2s}}  & =
I'_s & +\ \ds
\Delta' \frac{Q^{|2s|} (0) J_{-2s}(qR_{ex})}{1 + \tilde B_{2s}} X   \nn \\ && +\ \ds
\Delta \frac{Q^{|2s|} (0) J_{-2s}(qR_{ex})}{1 + \tilde B_{2s}} Y .
 \end{array}          \right.
         \ee                                       
 Here:
          \bea 
&&X = \sum \limits_s P^{|2s|} (0) J_{-2s}(qR_{ex}) I_s ,
       \\ &&
Y = \sum \limits_s Q^{|2s|} (0) J_{-2s}(qR_{ex}) I'_s .
        \eea
       We can find the quantities $ X $ and $ Y $ solving the system of equations:
        \be 
\left \{
\begin {array}    {l}
\displaystyle { X(1 + \alpha_q \Delta ) + Y \alpha_q \Delta' =
\sum \limits_s N_s \frac{P^{|2s|}(0)}{1 + \tilde A_{2s}} \,
J_{-2s} (qR_{ex}) }  ,    
     \\
\displaystyle { X \beta_q \Delta'  + Y (1 + \beta_q \Delta )=
\sum \limits_s N'_s \frac{Q^{|2s|}(0)}{1 + \tilde B_{2s}}  \,
J_{-2s} (qR_{ex}) }   .
\end{array}          \right.
        \ee
 Here
  \bea 
\alpha_q &=& \sum \limits_s \frac{A_{2s}^*}{1 + \tilde A_{2s}}
J_{-2s}^2 (qR_{ex}) , \nn\\
\beta_q &=& \sum \limits_s \frac{B_{2s}^*}{1 + \tilde B_{2s}}
J_{-2s}^2 (qR_{ex}) \nn \\
A_{2s}^* &=& [P^{|2s|} (0)]^2;
\quad B_{2s}^* = [Q^{|2s|} (0)]^2.
        \eea
 As a result we arrive at the expressions:
  \bea 
X &=& \frac{1 + \beta \Delta}{D}
\sum \limits_s N_s \frac{P^{|2s|}(0)}{1 + \tilde A_{2s}} \,
J_{-2s} (qR_{ex})  \nn\\&  -&
  \frac{\alpha \Delta'}{D}
\sum \limits_s N'_s \frac{Q^{|2s|}(0)}{1 + \tilde B_{2s}} \,
J_{-2s} (qR_{ex})   ,
  \nn\\\nn\\
Y& =& \frac{1 + \alpha \Delta}{D}
\sum \limits_s N'_s \frac{Q^{|2s|}(0)}{1 + \tilde B_{2s}} \,
J_{-2s} (qR_{ex})  \nn\\& -&
  \frac{\beta \Delta'}{D}
\sum \limits_s N_s \frac{P^{|2s|}(0)}{1 + \tilde A_{2s}} \,
J_{-2s} (qR_{ex})   ,
        \eea
 where the determinant of the system equals:
                 \be 
D = 1 + (\alpha_q + \beta_q) \Delta +
\alpha_q \beta_q (\Delta^2 - \Delta'^2) .
        \ee
 Sustituting the expressions (70) into the system (65), we can calculate the averages: 

       \bea 
I_s &=& \frac{N_s}{1 + \tilde A_{2s}} - \frac{\Delta +
\beta_q (\Delta^2 - \Delta'^2)}{D} \,
\frac{P^{|2s|}(0)}{1 + \tilde A_{2s}} \, J_{-2s} (qR_{ex}) 
  \nn\\ &\times&
\sum \limits_{s'} N_{s'} \frac{P^{|2s'|}(0)}{1 + \tilde A_{2s'}} \, J_{-2s'} (qR_{ex})
   -
\frac{\Delta'}{D} \,
\frac{P^{|2s|}(0)}{1 + \tilde A_{2s}} \,  
  \nn\\ &\times& J_{-2s} (qR_{ex})
\sum \limits_{s'} N'_{s'} \frac{Q^{|2s'|}(0)}{1 + \tilde B_{2s'}} \, J_{-2s'} (qR_{ex})   ,
       \\\nn\\
I'_s & =& \frac{N'_s}{1 + \tilde B_{2s}} - \frac{\Delta +
\alpha_q (\Delta^2 - \Delta'^2)}{D} \,
\frac{Q^{|2s|}(0)}{1 + \tilde A_{2s}} \,  J_{-2s} (qR_{ex}) 
   \nn \\&\times&
  \sum \limits_{s'} N'_{s'} \frac{Q^{|2s'|}(0)}{1 + \tilde A_{2s'}}J_{-2s'} (qR_{ex}) -
\frac{\Delta'}{D} \frac{Q^{|2s|}(0)}{1 + \tilde B_{2s}} \, 
 \nn\\&\times& J_{-2s} (qR_{ex}) \!
\sum \limits_{s'} N_{s'} \frac{P^{|2s'|}(0)}{1 + \tilde A_{2s'}}
J_{-2s'} (qR_{ex}).
        \eea

The asymptotic expressions for the quantities $ N_s $ and $ N'_s $ also can be found by the formula (24). For the FS with the unique extremal cross-section we have:
   \bea 
N_s & =& - \frac{m_\perp}{2 \pi^2 \hbar^3} \int dp_z P^0 (p_z) \delta_{s0}
- g_0 \Delta J_0 (qR_{ex}) \nn\\ 
&\times& P^{|2s|}(0) J_{-2s} (qR_{ex}) ,
\nn\\
N'_s &=& - g_0 \Delta' J_0 (qR_{ex}) Q^{|2s|}(0) J_{-2s} (qR_{ex}) .
        \eea
 Using the Eqs. (57) and (72)-(74), we obtain
  \bea 
&-& \sum_{\nu \nu'}
\frac{f_\nu - f_{\nu'}}{E_\nu - E_{\nu'}} \,
n_{\nu \nu'}^*{\bf (-q)}  n_{\nu' \nu} {\bf (q)} =
 g_0 \bigg [ 1 - \alpha_0 
  \nn\\&+&
 \frac{[(1 - \overline \alpha_0)^2
J_0^2(q R_{ex})][\Delta + \beta_q (\Delta^2 - \Delta'^2)]}
{1 + (\alpha_q + \beta_q)\Delta +
\alpha_q \beta_q (\Delta^2 - \Delta'^2)} \bigg].
        \eea
 Here we introduced additional notations:
                 \bea &&
\alpha_0 = \frac{A_0}{1 + \tilde A_0} \, ,\qquad
\overline \alpha_0 = \frac{\overline A_0}{1 + \tilde A_0} \, ,
 \nn\\ &&
A_0 =\left ( \frac{m_\perp}{2 \pi^2 \hbar^3 g_0}
\int P^0 (p_z) d p_z \right)^2 \, , \nn\\&&
\overline A_0 =  \frac{P^0 (0) m_\perp}{2 \pi^2 \hbar^3 g_0}
\int P^0 (p_z) d p_z .
        \eea

Passing to the limit $ q \to 0 $ in the Eq. (75) we obtain the final result for $ N_\zeta^*.$

\end{document}